\documentclass[aps, prb,twoside,twocolumn,english, floatfix, superscriptaddress]{revtex4-1}
\usepackage[T1]{fontenc}
\usepackage[utf8]{inputenc}
\usepackage{xcolor}
\usepackage{units}
\usepackage{babel}
\usepackage{textcomp}
\usepackage{amsmath}
\usepackage{graphicx}

\begin{document}
\title{An anomalous refraction of spin waves as a way to guide signals\\ in curved magnonic multimode waveguides}
\begin{abstract}
We present a method for efficient spin wave guiding within the magnonic nanostructures. Our technique is based on the anomalous refraction in the metamaterial flat slab.
The gradual change of the material parameters (saturation magnetization or magnetic anisotropy) across the slab allows tilting the wavefronts of the transmitted spin waves and controlling the refraction.
Numerical studies of the spin wave refraction are preceded by the analytical calculations of the phase shift acquired by the spin wave due to the change of material parameters in a confined area. 
We demonstrate that our findings can be used to guide the spin waves smoothly in curved waveguides, even through sharp bends, without reflection and scattering between different waveguide's modes, preserving the phase -- the quantity essential for wave computing.
\end{abstract}
\author{Szymon Mieszczak$\dagger$}
\email{szymon.mieszczak@amu.edu.pl}
\affiliation{Faculty of Physics, Adam Mickiewicz University, Pozna\'{n}, Uniwersytetu Poznanskiego 2, Pozna\'{n} 61-614, Poland}

\author{Oksana Busel$\dagger$}
% \email{opbusel@gmail.com}
\affiliation{National Technical University of Ukraine \textquotedblleft Igor Sikorsky
Kyiv Polytechnic Institute\textquotedblright , 37 Prosp. Peremohy,
Kyiv, 03056, Ukraine}
\author{Pawe\l{} Gruszecki}
\affiliation{Faculty of Physics, Adam Mickiewicz University, Pozna\'{n}, Uniwersytetu Poznanskiego 2, Pozna\'{n} 61-614, Poland}
\author{{Andriy N. Kuchko}}
\affiliation{National Technical University of Ukraine \textquotedblleft Igor Sikorsky
Kyiv Polytechnic Institute\textquotedblright , 37 Prosp. Peremohy,
Kyiv, 03056, Ukraine}
\affiliation{Institute of Magnetism of NAS of Ukraine, 36b Vernadskogo Avenue, Kyiv, 03142, Ukraine}
\author{Jaros\l aw W. K\l os}
\author{Maciej Krawczyk}
\affiliation{Faculty of Physics, Adam Mickiewicz University, Pozna\'{n}, Uniwersytetu Poznanskiego 2, Pozna\'{n} 61-614, Poland}

\maketitle

\section{Introduction}

The phase and amplitude are the fundamental characteristics of waves. The  processing of any kind of waves relies on the interference effects which depend on these characteristics. Thus, the control of spin waves' (SWs) phase and amplitude is essential in magnonics\cite{Demokritov13} to perform both analog\cite{Csaba2017,Khitun13} and digital\cite{Khitun2010} SW-based computing\cite{Chumak19}. 

One of the significant challenges limiting the application of SWs relates to the capability of coherent and weakly damped signal transmission. Fulfilling this condition is necessary to transmit the information, encoded in SW phase or amplitude, between particular parts of a magnonic circuit, sometimes in a grid of interconnected and crossed waveguides, enabling a flow of SWs in different directions\cite{Khitun2010}. Typically, the interconnections are realized by waveguides being narrow and flat ferromagnetic stripes. 
Here arises the problem of the SW scattering on bends of waveguides. If the static magnetization is saturated and oriented along the direction of the external magnetic field, then the magnetic surface charges, generated by the normal to the surface component of the magnetization, will change at the bends of the waveguide. 
On the other hand, if the external magnetic field is low\cite{vogt2012spin} or the waveguide is properly patterned\cite{haldar2016reconfigurable},  the static magnetization follows the shape of the curved waveguide due to the shape anisotropy.
Nevertheless, the magnetic volume charges will be generated due to curvilinear magnetic configuration. 
In addition, the exchange interaction will be modified, which induces the effects equivalent to the presence of the anisotropy field or the field of Dzyaloshinskii–Moriya interaction\cite{Sheka19,Gaididei_2017,Tkachenko12}.

The SWs' wavelength is a few orders of magnitude shorter than the electromagnetic waves of corresponding frequencies\cite{Csaba2017}. Therefore,  typical magnonic waveguides of the width accessible in photo-lithographic fabrication techniques are multi-modal waveguides even in the GHz-frequency range. 
The fabrication of a single-mode waveguide for SWs is difficult since their widths would need to be narrow, especially for high-frequency SWs\cite{Wang2019}. 
Up-to-date, the single-mode waveguides for short SWs can be realized in the systems utilizing domain walls as magnonic waveguides, since they can create narrow potential wells where SW modes can be localized and propagate lengthwise \cite{garcia2015,wagner2016,banerjee2017,lan2015spin,henry2019,gruszecki2019chapter}. 

Difficulties associated with SW wavelengths dependence on the direction of propagation (anisotropic dispersion relation) are negligible for short-wavelength SWs where exchange interactions of isotropic nature dominate over the anisotropic dipolar interactions. 
Also, for the magnetic configuration where the magnetic field is applied perpendicularly to the film's surface, the SW dynamics is naturally isotropic, independently on the frequency of SW. The only obstacle related to that geometry is a high bias field demanded to magnetize the sample uniformly. It can be overcome in materials with strong out-of-plane anisotropy, but these are usually characterized by high SW damping.

In multi-mode waveguides, the mechanism, which leads to the decoherence of the propagating SWs is scattering to other, perpendicularly quantized modes. 
Therefore, the signal loses the information encoded in the phase.
Another consequence is that the SW propagates along a longer zig-zag shaped path\cite{Clausen11}, which can also be interpreted as a redistribution of the momentum (wave vector) between the components, that are transferal and longitudinal to the waveguide's axis. It is worth to note, that the transverse quantization of the modes in the planar waveguide is related not only to the width of the structure but can be also introduced additionally by the periodic patterning along the waveguide\cite{Klos14,Pan17,Lee2009, Ciubotaru_2012}.

We pointed out that the scattering between the modes is one of the most important factors for the SW decoherence at the bends of the magnonic waveguide. Therefore, the question arises: can we modify the properties of the bending region to block the redistribution of incoming mode into the different outgoing modes, keeping the transmission as high as possible?
One possible solution is to fill the bending region by the material of spatially tailored properties, which will refract the SW and redirect its propagation strictly along the outgoing section of the waveguide.
In other words, we should look for the particular kind of so-called graded index (GRIN) element for SWs \cite{toedt2016design,Vogel2019,Gruszecki2018,Whitehead2018,whitehead2019GRINlens,gieniusz2017switching,dzyapko2016heatLens,Vogel2019}.
Recently GRIN elements have been used to bend SWs for in-plane\cite{Vogel2019} and out-of-plane\cite{whitehead2019GRINlens} magnetized films.
Another exciting idea is an application of SW lenses, in particular flat metalenses \cite{zelent2019spin}, i.e., lenses of fixed width introducing different phase delay of transmitted waves alongside the interface. 
Except for the changes in the phases of refracted waves, the GRIN element should not introduce significant changes to their amplitudes. 
The interplay between the material parameters of the slab and its sizes determines the conditions for the resonanant transmission\cite{Klos2018,Dobrovolskiy19}. Therefore, we need to design a system to work in the conditions close to the resonant  transmission.

In this paper, we employ  the anomalous refraction\cite{yu2011light} achieved in the GRIN slab to change the direction of coherently propagating SWs at the bend of the waveguide. For anomalous refraction, the wavefronts of refracted waves are tilted at a desirable angle with respect to wavefronts of the incident waves, even at normal incidence. 
This phenomenon requires a linear change of the phase of the transmitted waves alongside the interface, where the refraction takes place. Its description requires the generalization of the Snell's law \cite{yu2011light, Mulkers2018}. To the best of our knowledge, this effect has not yet been exploited for guiding SWs in the waveguides.

We develop the analytical theory for the scattering of exchange SWs on the homogeneous ferromagnetic slab of finite width embedded in a ferromagnetic layer. Minimizing the total energy, we derive the boundary conditions on the interfaces between the slab and its surroundings. We obtain the complete relations between the phases and amplitudes of the incident and scattered SWs. These calculations are successfully compared to micromagnetic simulations.
Then, we use our findings to demonstrate both analytically and numerically an anomalous refraction for the purely exchange SWs incident from a waveguide to a semi-infinite film through a flat magnonic GRIN slab. We treat the GRIN element as an inhomogeneous slab linking the input and output branches of the waveguide at the bend. 

The manuscript is organized as follows. In the next section, we describe the analytical model. In the Sec. III we show and discuss the results of the analytical and numerical studies, which are summarized in the Sec. IV. In Appendix A we present details of anyltical calculation, while in B details of micromagnetic simulations.
\bigbreak

\section{Model and methods}

\subsection{Boundary conditions problem}

\begin{figure}
\includegraphics[width=1\columnwidth]{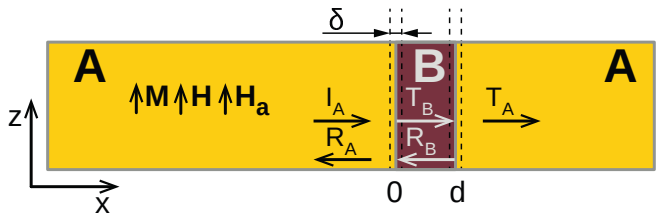}
\caption{
The SWs scattering on a ferromagnetic layer (dark area) embedded in a ferromagnetic matrix (light area).
The uniaxial anisotropy field $\mathbf{H}_{a}$ and the static magnetization $\mathbf{M}$ for ferromagnetic matrix (A) and layer (B) are parallel to each other and tangential to the plane of the layer. The layer B of the thickness $d$ is exchange coupled to the matrix A by the interfaces of the thickness $\delta$.}\label{fig: geometry of the system} 
\end{figure}

Let us consider SWs propagating (along the \emph{x}-axis) through
a ferromagnetic layer B (\emph{$0<x<d$}) embedded as a slab between two half-spaces of the ferromagnetic matrix A (\emph{$x\leq0$} and \emph{$x\geq d$}), 
as shown in Fig.~\ref{fig: geometry of the system}.
The slab B is exchange coupled by thin interfaces of the thickness $\delta$ to the matrix A. For simplicity's sake, we assumed that the system is uniform and infinitely extended in the $x-z$ plane.
We consider the case when the static magnetizations $\mathbf{M}_{\mathrm{A}}$
and $\mathbf{M}_{\mathrm{B}}$ are oriented along the $z$-axis (see Fig.~\ref{fig: geometry of the system}) and are parallel to each other everywhere in the system: $\mathbf{M}_{\mathrm{A}\left(\mathrm{B}\right)}=\left[0,0,M_{S,\mathrm{A}\left(\mathrm{B}\right)}\right]$, where $M_{S,\mathrm{A}\left(\mathrm{B}\right)}$ denotes the saturation magnetization. The indexes A and B denote materials of the matrix and the slab, respectively.

The dynamics of magnetization in an effective magnetic field can be described by the Landau-Lifshitz equation (LLE):
\begin{equation}
\frac{\partial\mathbf{M}_{\mathrm{A}\left(\mathrm{B}\right)}}{\partial t}=-\mu_{0}|\gamma|\left(\mathbf{M}_{\mathrm{A}\left(\mathrm{B}\right)}\times\mathbf{H}_{{\rm eff},\mathrm{A}\left(\mathrm{B}\right)}\right),
\label{Eq:LLE}
\end{equation}
where the effective field is a variational derivative of the energy density $w$ with respect to the magnetization vector:
 $\mathbf{H}_{{\rm eff},\mathrm{A}\left(\mathrm{B}\right)}=-({1}/{\mu_0})\delta w/\delta\mathbf{M}_{\mathrm{A}\left(\mathrm{B}\right)}$.
The parameter $\gamma$ is the gyromagnetic ratio and $\mu_{0}$ is the permeability
of the vacuum. The total magnetic energy density of
the system $w$ includes: the density of the Zeeman energy $\left(-\mu_0\mathbf{H}\cdot\mathbf{M}_{\mathrm{A}\left(\mathrm{B}\right)}\right)$,
with the external magnetic field $\mathbf{H}$; the exchange energy density
$\left(1/2\alpha_{\mathrm{A}\left(\mathrm{B}\right)}\left(\partial\mathbf{M}_{\mathrm{A}\left(\mathrm{B}\right)}/\partial x_{i}\right)^{2}\right)$,
with the exchange interaction parameter $\alpha_{\mathrm{A}\left(\mathrm{B}\right)}=2A_{{\rm ex},\rm{A}\left(\mathrm{B}\right)}/M_{S,\mathrm{A}\left(\mathrm{B}\right)}^{2}$,
where $A_{{\rm ex},A\left(\mathrm{B}\right)}$ is the exchange stiffness constant
in the material A or B respectively; the density of anisotropy energy $\left(-1/2\beta_{\mathrm{A}\left(\mathrm{B}\right)}\left(\mathbf{M}_{\mathrm{A}\left(\mathrm{B}\right)}\cdot\mathbf{n}_{a}\right)^{2}\right)$,
where $\beta_{\mathrm{A}\left(\mathrm{B}\right)}=2K_{\mathrm{A}\left(\mathrm{B}\right)}/M_{S,\mathrm{A}\left(\mathrm{B}\right)}^{2}$
and $\mathbf{n}_{a}$ is the unit vector of the easy axis. Anisotropy energy density is expressed by the uniaxial
anisotropy constant $K_{\mathrm{A}\left(\mathrm{B}\right)}$. 
Assuming that the thickness of the interface $\delta$ is smaller than the exchange length $\lambda_{\rm ex}=\sqrt{2 A_{\rm ex}/\mu_0 M_{S}^2}$, we can neglect the structure of the interfaces (roughness, material mixing) and introduce the coupling parameters $A$ , which are the parameter of the interlayer exchanges and can be expressed via the interface thickness $\delta$\cite{Cochran92,Kruglyak2014}.
We postulate the exchange type of coupling characterized by the energy density at the interfaces: $x={0,d}$ (i.e. the energy per unit area): $-A\mathbf{M}_{\mathrm{A}}\cdot\mathbf{M}_{\mathrm{B}}$, where  $A=\left(A_{{\rm ex},\mathrm{A}}+A_{{\rm ex},\mathrm{B}}\right)/(2M_{S,\mathrm{A}}M_{S,\mathrm{B}}\delta)$ is coupling parameter.

Minimizing the total energy, we can derive boundary conditions:
\begin{equation}
\begin{cases}
\begin{array}{c}
\left(\alpha_{\mathrm{A}}\frac{\partial}{\partial x}+A\frac{M_{S,B}}{M_{S,A}}\right)m_{\mathrm{A}}-Am_{\mathrm{B}}=0\\
\left(\alpha_{\mathrm{B}}\frac{\partial}{\partial x}-A\frac{M_{S,A}}{M_{S,B}}\right)m_{\mathrm{B}}+Am_{\mathrm{A}}=0
\end{array}, & x=0\end{cases},\label{eq:bound.cond.1}
\end{equation}
\begin{equation}
\begin{cases}
\begin{array}{c}
\left(\alpha_{\mathrm{B}}\frac{\partial}{\partial x}+A\frac{M_{S,A}}{M_{S,B}}\right)m_{\mathrm{B}}-Am_{A}=0\\
\left(\alpha_{\mathrm{A}}\frac{\partial}{\partial x}-A\frac{M_{S,B}}{M_{S,A}}\right)m_{\mathrm{A}}+Am_{\mathrm{B}}=0
\end{array}, & x=d\end{cases}.\label{eq:bound.cond.2}
\end{equation}
At each interface between matrix and slab, the solutions of the LLE satisfy the boundary conditions for the amplitudes of the dynamical components of magnetization
 $\mathbf{m}_{\mathrm{A}\left(\mathrm{B}\right)}=\left[m_{\mathrm{A}\left(\mathrm{B}\right),x},
 m_{\mathrm{A}\left(\mathrm{B}\right),y},0\right]$
(for convenience, in Eqs.~(2) and (3) they are expressed via the cyclic variables  $m_{\mathrm{A}\left(\mathrm{B}\right)}\mathbf{=}m_{\mathrm{A}\left(\mathrm{B}\right),x}\pm im_{\mathrm{A}\left(\mathrm{B}\right),y}$,
where $i$ is the imaginary unit) for every interface, namely at $x=0$ and  $x=d$.

We are looking for a solution in linear regime, i.e., describing the harmonic precession with the angular frequency $\omega$: $\mathbf{m}\left(\mathbf{r},t\right)=\mathbf{m}\left(\mathbf{r}\right)\exp\left(i\omega t\right)$.

For the general stationary solution $\bf{m}(\mathbf{r})$,  we have to include the waves propagating both to the left and to the right. This can be mathematically expressed as:

\begin{equation}
\begin{array}{cc}
m_{\mathrm{A}}=I_{\mathrm{A}}e^{ik_{\mathrm{A}}x}+r_{\mathrm{A}}e^{-ik_{\mathrm{A}}x}, & x\leq0\end{array},\label{eq:mA1}
\end{equation}
\begin{equation}
\begin{array}{cc}
m_{\mathrm{B}}=t_{\mathrm{B}}e^{ik_{\mathrm{B}}x}+r_{\mathrm{B}}e^{-ik_{\mathrm{B}}x}, & 0<x<d\end{array},\label{eq:mB}
\end{equation}
\begin{equation}
\begin{array}{cc}
m_{\mathrm{A}}=t_{\mathrm{A}}e^{ik_{\mathrm{A}}x}, & x\geq d\end{array}.\label{eq:mA2}
\end{equation}
The parameters $k_{\rm{A}}$ and $k_{\rm{B}}$ stand for the wave numbers in the matrix and the slab, respectively. The wave numbers $k_{\rm{A}}$ and $k_{\rm{B}}$ depend on the external magnetic field $H$ and material parameters: magnetization saturation $M_{S,\rm{A}(\rm{B})}$ and anisotropy field $H_{\rm{a},\mathrm{A(B)}}=2K_{\mathrm{A(B)}}/(\mu_0 M_{S,\rm{A(B)}})$.
The amplitude of incoming wave is normalized to  one: $I_{\mathrm{A}}=1$. All the remaining amplitudes $r_{\mathrm{A}}$, $t_{\mathrm{B}}$,
$r_{\mathrm{B}}$, $t_{\mathrm{A}}$ are, in general, complex valued: $R_{\mathrm{A}}e^{i\varphi_{R_{\mathrm{A}}}}$,
$T_{\mathrm{A}}e^{i\varphi_{T_{\mathrm{A}}}}$, $T_{\mathrm{B}}e^{i\varphi_{T_{\mathrm{B}}}}$, $R_{\mathrm{B}}e^{i\varphi_{R_{\mathrm{B}}}}$ and contain the information about the real amplitudes $R_{\mathrm{A}}$, $T_{\mathrm{B}}$, $R_{\mathrm{B}}$, $T_{\mathrm{A}}$
and phases $\varphi_{R_{\mathrm{A}}}$, $\varphi_{T_{\mathrm{A}}}$, $\varphi_{T_{\mathrm{B}}}$,
$\varphi_{R_{\mathrm{B}}}$.

For calculating the complex amplitudes of SWs, namely  $r_{\mathrm{A}}$, $t_{\mathrm{B}}$, $r_{\mathrm{B}}$ and $t_{\mathrm{A}}$ (presented in Appendix~A, Eqs.~(11)--(14)), we use the boundary conditions Eqs.~(\ref{eq:bound.cond.1}) and (\ref{eq:bound.cond.2}).

\subsection{Huygens--Fresnel principle \\and generalized Snell's law}

Let us discuss the design of a GRIN slab in the form of the rectangular region with a gradual change of the magnetic parameters enabling steer the direction of the transmitted waves. To describe the wave refraction in this system, we use the Huygens--Fresnel principle\cite{Hecht2017}. This concept was developed for optics but can be adapted to other types of  waves\cite{Tang15}.
The postulation states that every point on a wavefront is itself the source of the cylindrical wavelets, and the sum of these cylindrical wavelets forms the new wavefront.
In the considered case, we assume that the source points are located at the right interface of the GRIN slab (at $x=d$) and are aligned along the $y$-axis. 
We can calculate the complex amplitude at any position on the right side of the GRIN slab using the amplitude and phase shift obtained from the solution of the boundary problem. 
Postulation given by Huygens--Fresnel can be mathematically expressed as:
\begin{equation}
\mathcal{T}\left(\mathbf{r}\right)=\sum_{j}\frac{T_{j}e^{i[\mathbf{k}\cdot(\mathbf{r} - \mathbf{r}_j)+ \phi_{j}]}}{\left|\mathbf{r}-\mathbf{r}_j\right|}.\label{eq: hug}
\end{equation}
The summation is done over the interface of the GRIN slab (at the position $x=d$) at large number of locations $\mathbf{r}_j=[d,y_j]$. The phase $\phi_j$ is tailored by changing the saturation magnetization or anisotropy field in the GRIN slab.
The partial amplitude $T_j$ indicates the transmittance of the SW through the GRIN slab into the semi-infinite medium A defined as $\mathrm{abs}(t_{\mathrm{A}})^{2}$ 
and the phase $\phi_j$ defined as $\mathrm{arg}(t_{\mathrm{A}})$ (Eq.~(\ref{eq:t_A}) in Appendix A).

Another approach for calculating the angle of the refracted wave is to use the generalized Snell law\cite{yu2011light, Yu2014}. It states that a nonzero gradient of phase along the interface induces an additional perpendicular to the interface component of the  wavevector. Mathematically it can be expressed by the following formula:
\begin{equation}
    k_{\text{incident},y}=k_{\text{refracted},y}+\frac{1}{\lambda}\frac{d\phi}{dy}, \label{eq: snell}
\end{equation}
where $\phi$ is induced phase along the $y$-axis and $k_\mathrm{incident}$ and $k_\mathrm{refracted}$ are the wave numbers of the incident and refracted wave, respectively.
 In the case with known $\phi(y)$ dependence along the $y$-axis, the generalized Snell law together with SW dispersion relation, can be used to predict the overall direction of the wavevector of the outgoing SWs from the GRIN slab. In our case, for normal incident waves, $k_{\mathrm{incident},y}=0$, therefore, $k_{\text{refracted},y}=-(1/{\lambda}) d{\phi}/dy$. The missing component of wavevector, perpendicular to the interface, can be calculated with the use of SW dispersion relation $k(\omega)$: $k^2_{\text{refracted},x}=k^2(\omega)-k^2_{\text{refracted},y}$.  Overall, the GRIN slab acts here as a metasurface. Note, that in photonics, metasurfaces can also have widths comparable to the wavelength\cite{Jang2018}.

\begin{figure}
\includegraphics[width=1\columnwidth]{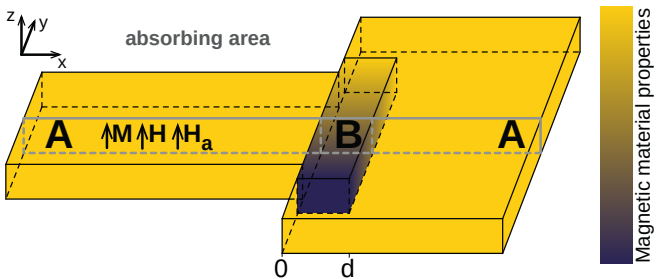}\caption{\label{fig: geometry_of_the_system_2} The geometry of the simulation system with the GRIN element. The system consists of the ferromagnetic slab (B) characterized by the gradient of magnetic parameters (e.g., saturation magnetization $M_S$ or anisotropy field $H_\text{a}$) which links the straight section of the waveguide and the semi-infinite plane made of the homogeneous material A. The magnetic parameters in the slab B are changing in the $y$-direction. At each $x$--$z$ cross-section (see Fig.~\ref{fig: geometry of the system}), the phase shift of transmitted wave is different, which allows refracting the plane wave propagating initially in the $x$-direction. The absorbing material is placed at the sides of the simulated system, in order to avoid the impact of boundaries. The external magnetic field $\mathbf{H}$ is applied along the $z$-direction; the static magnetization and the anisotropy field are aligned with the external field.
}
\end{figure}

\section{Results}

In order to design the GRIN slab, that is schematically shown in Fig.~\ref{fig: geometry_of_the_system_2}, we propose to use CoFeB layer as a base material due to high magnetization saturation, relatively low SW damping\cite{Conca13,Kuswik2017}, which is crucial for SW processing,\cite{Chumak_2017} and because of the possibility of tailoring the material parameters, required to obtain anomalous refraction. 

By implantation of Ga ions, we can locally reduce the magnetization and create the distribution of magnetization saturation of high spatial resolution\cite{Maziewski12,Wawro18,Dobrovolskiy19}. From the other side, when the sufficiently  thin layer of the CoFeB is deposited on the MgO, the surface-induced out-of-plane anisotropy can force the perpendicular orientation of the magnetization\cite{ikeda_2010,Klos2018}. The effective anisotropy field (including both the shape and surface, magnetocrystalline anisotropy) can be controlled by the annealing in the fabrication process or by the application of the electric field\cite{Naik2014,Miwa_2018,Rana19}. The perpendicular orientation of the magnetization ensures the isotropic SW dispersion, which simplifies the design of refraction effects.

At the initial stage, we will perform the analytical studies of the SWs transmission through the slab formed in a CoFeB matrix by the modification of the magnetization saturation or the anisotropy field. 

The particular attention will be paid to the identification of the resonances of the slab, where the amplitude of transmitted SWs is the highest. 
The results will be crosschecked by micromagnetic simulations. 
Then, we will present the outcomes of micromagnetic simulations for the guiding of the SWs in the waveguide with the GRIN slab at the bend. The GRIN slab will be designed using the results form the initial stage and used to guide the SWs coherently through the bend of the magnonic waveguide.

\subsection{Spin wave propagation through the slab -- one dimensional scenario}

Let us analyze the SW transmission through a uniform slab of finite width with modified either saturation magnetization or uniaxial anisotropy field. Eqs.~(\ref{eq:mA1})--(\ref{eq:mA2}), with Eqs.~(\ref{eq:t_A})--(\ref{eq:k_A(B)}) are taken into account to determine the phase shift $\mathrm{arg}(t_{\mathrm{A}})$ and the transmittance $\mathrm{abs}(t_{\mathrm{A}})^{2}$ of the transmitted SWs in dependence on frequency or other considered magnetic parameters (e.g., the saturation magnetization or the anisotropy field).

To demonstrate various aspects of the system, as well as to check the validity of the model, we will show the results of the analytical calculations and their comparison with the results of the micromagnetic simulations for four scenarios: (i) $M_\mathrm{S,B}=800$ kA/m in the slab is fixed, the exchange constant $A_{\mathrm{ex,B}}$ is equal to 20 pJ/m, the frequency of SWs varies in the range of 14-40 GHz and the uniaxial anisotropy is neglected, (ii) $M_\mathrm{S,B}=1200$ kA/m, $A_{\mathrm{ex,B}}= 27$ pJ/m, frequency of SWs varies in the range of 14-40 GHz and the uniaxial anisotropy $K_\mathrm{B}=50$ kJ/m$^3$ is included, (iii) $M_\mathrm{S,B}$ varies in the slab in the range of 300-800~kA/m,  $A_{\mathrm{ex,B}}=20$ pJ/m, frequency of SWs is 25 GHz and the uniaxial anisotropy is neglected, (iv) $M_\mathrm{S,B}=1200$ kA/m, $A_{\mathrm{ex,B}}=27$ pJ/m, SWs frequency is 25 GHz and the uniaxial anisotropy $K_{\mathrm{B}}$ changes in the range of 0-490 kJ/m$^3$. In all cases, the slab is 150 nm wide.
Surrounding material A is assumed to be made from CoFeB with $M_\mathrm{S,A}=1200$ kA/m, $A_{\mathrm{ex,A}}=27$ pJ/m and $K_\mathrm{\mathrm{A}}=0$. The external magnetic field $\mu_0 H=0.5$ T is aligned along the $z$-axis. Details of the micromagnetic simulations are presented in Appendix B.

\begin{figure}
\includegraphics[width=1\columnwidth]{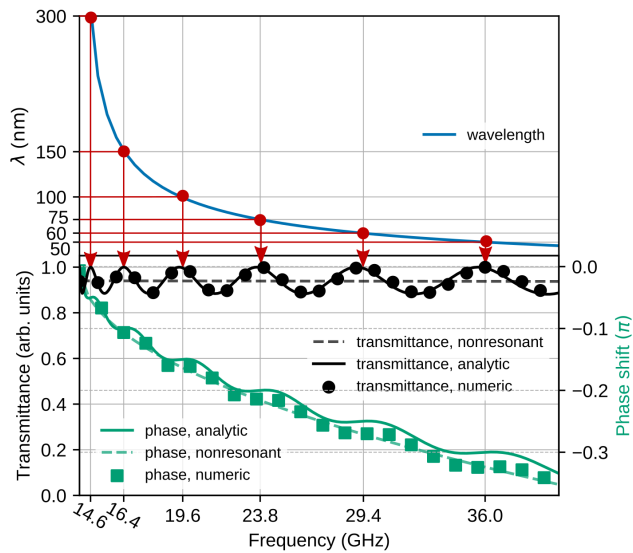}

\caption{\label{fig:freqms} The SWs' wavelength (blue color) as the function of frequency. The red dots indicate fulfilled resonant conditions by Eq.~(\ref{eq: resonance}).
Enhancement of transmittance at these frequencies can be observed. 
The values on the horizontal axis represent the resonant frequencies. The transmittance (black color) and the phase shift (green color) for SWs traveling through the
$150$~nm-wide slab with respect to the frequency. The dots and the squares represent
the values obtained in the numerical simulations, while the solid
lines represent the analytical results. The dashed lines represent the case when the reflection in the system was neglected. The value of the external field is equal to $0.5$~T,
reduced $M_{\mathrm{S}}$ in the slab is equal to $800$~kA/m and reduced $A_{\mathrm{ex}}$ to
$20$~pJ/m. } 
\end{figure}

\begin{figure}
\includegraphics[width=1\columnwidth]{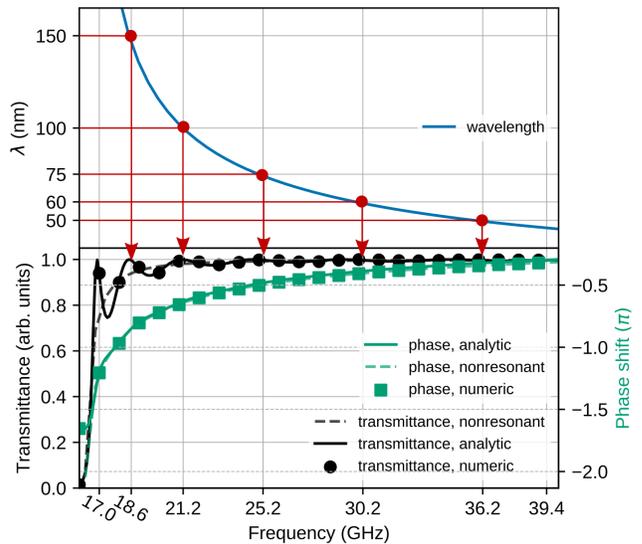}

\caption{\label{fig: freqani} The SWs' wavelength (blue color) as the function of frequency. The red dots indicate fulfilled resonant conditions, Eq.~(\ref{eq: resonance}), where we observe an enhancement of the transmittance. 
The values on the horizontal axis represent the resonant conditions. Transmittance (black color) and phase shift (green color) for SWs propagating
through the $150$~nm-wide slab with respect to the frequency. The dots and the squares
represent the values obtained in numerical simulation, while the solid
lines represent the analytical results. The dashed lines represent the case when the reflection in the system is neglected. The value of the external field is equal to $0.5$~T and the uniaxial anisotropy constant $K=5$~kJ/m$^{3}$ within
the slab.}
\end{figure}

We are going now to analyze the dependence of the transmittance and phase shift of transmitted SWs on the frequency for the slab formed by the modification of saturation magnetization and uniaxial anisotropy, cases (i) and (ii). The results for these two cases are presented in Fig. \ref{fig:freqms} and \ref{fig: freqani}, respectively.
To explain these frequency dependencies, we should discuss the role of the transmission of exchange SWs on the SWs dispersion relation
\begin{equation}
    k(\omega)=\sqrt{a\left(\frac{\omega}{\omega_0}-b\right)},\label{eq:k}
\end{equation}
where  $\omega_0=\gamma\mu_0 H$ is proportional to the value of external field $H$ and is expressed in  the angular frequency unit, the factor $a=M_{\rm S}\mu_0 H/(2A_{\rm ex})$ is proportional to the saturation magnetization, and the term $b=1+H_{\rm a}/H$ changes linearly with the anisotropy field $H_{\rm a}=2K/(\mu_0 M_{\rm S})$. By the change of  $M_\mathrm{S}$,  the wavevector is scaled, regardless of the range of frequencies. However, the impact of the $H_{\rm a}$ on the wavevector is significant only for small frequencies ($\omega$ is similar to $\omega_0$) when the additive term $b$ cannot be neglected. Moreover, for lower frequencies and  sufficiently large  positive value of the uniaxial anisotropy ($b>\omega/\omega_0$), the wave vector becomes imaginary and the SWs can only tunnel.

The decrease of $M_{\rm S}$ in the slab B (for $A_{\rm ex}$, $H$, $K$ kept constant in the system) results in the decrease of $k$ and the reduction of the phase acquired by the transmitted SW. As a result, the phase shift will be negative (referring to the SWs propagating at the same distance $d$ in the matrix A where $M_{\mathrm{S}}$ was not reduced). This negative phase shift is growing with the increasing frequency $\Delta\varphi(\omega)=d \left[k(\omega,M_{\mathrm{S,B}})-k(\omega,M_{\mathrm{S,A}})\right]$ --  see green lines and points in Fig.~\ref{fig:freqms}, because $k(\omega)$ is an increasing function of the frequency. The increase of the anisotropy field $H_\mathrm{a}$ in the slab B reduces the value of $k$. Therefore, the SWs gain additional phase during the transmission through the slab (referring to the SWs propagating at the same distance in the matrix A where $H_\mathrm{a}$ were not added). This positive phase shift is growing (see green lines and points in Fig.~\ref{fig: freqani}) for the same reasons as in the case of the slab formed by the change of $M_\mathrm{S}$.

It is visible that the value of transmittance oscillates. It is especially noticeable in the case of the slab induced by the $M_\mathrm{S}$ reduction, Fig.~\ref{fig:freqms}. To explain this behavior, the dispersion relations of SWs as the dependencies of wavelength on frequency $\lambda(\omega)=2\pi/k(\omega)$ are plotted in Fig.~\ref{fig:freqms} and Fig.~\ref{fig: freqani} (cf. Eq.~(\ref{eq:k})). The resonance condition for wave transmitted through the slab of the width $d$ reads:
\begin{equation}
2d=N\lambda(\omega),\label{eq: resonance}
\end{equation}
where $N$ is the natural number. 
The condition Eq.~(\ref{eq: resonance}) corresponds to the constructive interference of the wave after the round trip at a distance of $2d$. The frequencies for these resonances are marked in Fig.~\ref{fig:freqms} and Fig.~\ref{fig: freqani} by the red arrows. 
They match the locations of the maxima of the transmittance.
In order to confirm the presence of standing resonance modes within the slab, we performed both the analytical calculations and micromagnetic simulations.
We calculated the squared dynamical magnetization, averaged in time, at the frequencies corresponding to the maxima of the transmission. It can be observed that the slab works as a resonator (see the results in Appendix A), similar to Fabry-Perot resonators known in optics\cite{Hecht2017}. 

One can predict the phase shift, which SWs gain during the transmission through the slab, by defining a magnonic refractive index\cite{Vogel2019,Stigloher2016}. However, this approach takes into account only the refractive properties of the bulk material expressed in the dispersion relation. By solving the LLE (Eq.~(\ref{Eq:LLE})) with clearly defined boundary conditions (Eqs.~(\ref{eq:bound.cond.1}) and (\ref{eq:bound.cond.2})), we can obtain full information about the SWs' refraction in the system. Hence, we can see in Figs.~\ref{fig:freqms} and \ref{fig: freqani} the corrections, which come from taking into account the reflection from boundaries. 
The dashed lines with a lighter color indicate the cases when no reflection is considered in the system. As we can see, even for strong exchange coupling (when the majority of energy is transmitted for any condition) at the interface, $A$, the difference is noticeable. For weaker coupling, where a more significant fraction of energy would be turned back, the impact on the phase shift and the transmittance could be even more significant\cite{Klos2018}. This points out that the design of the GRIN slab has to take into account the presence of the resonances.

Numerical calculations fully support the analytical approach presented in this part, thus the validated model will be used for further stages of this study.

Let us discuss now the cases (iii) and (iv), described at the beginning of the section. Here, we aim to control the SW guiding in the confined structures like waveguides using GRIN slab, so in the following we introduce the change of the material parameters, saturation magnetization $M_\mathrm{S}$ and the anisotropy field $H_\mathrm{a}$. According to Fig.~\ref{fig: geometry_of_the_system_2}, we will change the material parameters of the slab at fixed frequency of SWs  $f=\omega/(2\pi)=25$ GHz. 

\begin{figure}
\includegraphics[width=1\columnwidth]{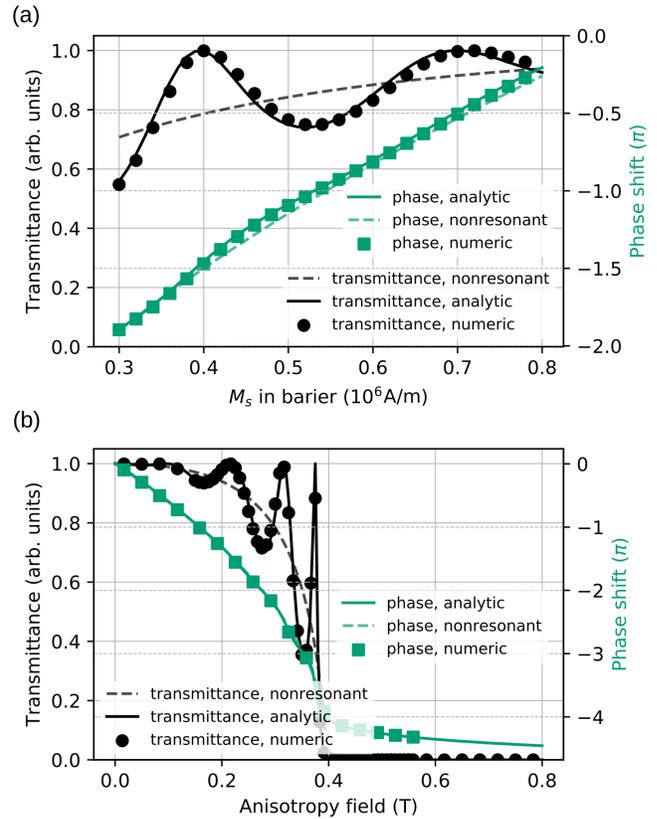}
\caption{\label{fig: msani}The transmittance (black color) and the phase shift (green
color) for SWs traveling through the $150$~nm-wide slab with respect to
(a) $M_{\mathrm{S}}$ and (b) the anisotropy field in the slab. The dots and the squares represent the values obtained in numerical simulation, while the solid lines represent the analytical results. The dashed lines represent the nonresonant case. The constant frequency equal to $25$~GHz and external field equal to $0.5$~T are considered.
(a) The reduced exchange $A_{\rm ex}$ is equal to $20$~pJ/m.  Resonant peaks are visible for the specific values of $M_{\mathrm{S}}$.
(b) $M_\mathrm{S}$ and $A_{\rm ex}$ remain the same as in the matrix, i.e., $1200$~kA/m and $27$~pJ/m, respectively.}
\end{figure}

In Fig.~\ref{fig: msani}(a), the transmittance and the phase shift with respect to $M_{\mathrm{S}}$ in the slab are presented. The resonant phenomenon is visible, like in the previous cases. The modulation of the transmittance is around 20\%.  The dashed black line represents the transmittance of SWs through the slab when the reflection is neglected. The impact of the resonances is visible, and taking into account the resonant effect is well-based. The phase shift changes rather smoothly in the considered range of $M_{\mathrm{S}}$, and in total, the difference is around $1.75\pi$. Deviations of the green solid and squared lines (analytical and numerical results, respectively) from the dashed line are minor and visible mainly near the resonances. In Fig.~\ref{fig: msani}(b), there are the results for the transmittance and the phase shift as function of the anisotropy constant $K$ in the slab in the range of 0-490~kJ/m$^3$, which is equivalent to the presence of the anisotropy field $\mu_0 H_\mathrm{a}$ in the range 0-0.8~T.

We can see that for the anisotropy field larger than 0.4~T, the transmittance drops dramatically, and the resonance peaks are not observed anymore. This feature is attributed to the lack of oscillating solution within the slab. The SWs need to tunnel through the slab with a significant reduction of the amplitude.
Just below this value ($H_\mathrm{a}=2\pi/(\gamma \mu_0) f - H$), the resonant behavior of the transmittance is observed with a high variation of the transmittance.

\subsection{Anomalous refraction of spin waves}

Based on the model developed in Section II, we will design a GRIN slab that bends wavefronts of the incident waves in the desired way. The slab is an element that modifies the phase of the incoming plane waves (at the interface located in $x=0$) to gain the linear change of the phase of the transmitted waves alongside the interface (at $x=d$).
This idea is based on the general concept of tailoring the phase changes alongside the interface and provide the desired functionality, like focusing\cite{zelent2019spin}, beam steering or delay \cite{Tian2019}. 
In the present case, we want to design an element suitable for the change of the direction of SWs propagation. 
Let us study numerically the case of the two-dimensional slab  with the gradient of $M_\mathrm{S}$, like it is schematically presented in Fig.~\ref{fig: geometry_of_the_system_2}. 
In the section of the straight and flat waveguide of the width $100$~nm, we excite the plane wave, which propagates along the $x$-axis. The waveguide is attached to the large plate of the same thickness and made of the same material. At the front of the plate, we defined the $150$~nm-wide slab -- see Fig.~\ref{fig: geometry_of_the_system_2}. In the slab, we take the reduced value of $M_\mathrm{S}$, which is increasing from 300 kA/m up to 800 kA/m on the distance $100$~nm, in direct contact with the waveguide. Above $M_\mathrm{S}$ goes to the value of a semi-infinite plane, it is $M_{S}=1200$~kA/m.

\begin{figure}
\includegraphics[width=1\columnwidth]{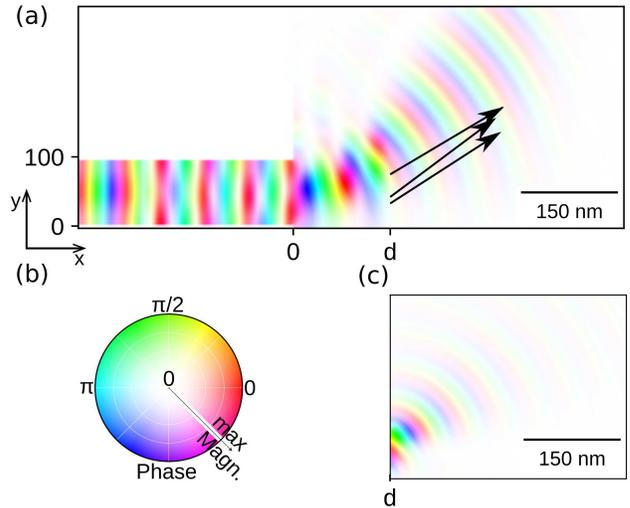}\caption{\label{fig: anomal }  
The squared amplitude of SWs propagating in the system shown in Fig.~\ref{fig: geometry_of_the_system_2},  at the frequency 25 GHz. (a) When SWs reach the semi-infinite medium after the slab, the new wavefront starts to form due to the gradient of magnetic parameters.
The bending of SWs reaches the angle of ca. $36{^\circ}$. (b) The color palette of SW. The color indicates the phase, while the intensity indicates the amplitude. (c) The bending of SWs ca. $36^{\circ}$ obtained according to Huygens--Fresnel formula (Eq.~(\ref{eq: hug})). The source points of cylindrical SWs are located on the right border of the slab -- at the position $d$. The amplitude of the transmittance and the phase shift are obtained from the boundary conditions problem, like in Fig.~\ref{fig: msani}(a).
}
\end{figure}

The results of the numerical simulations are presented in Fig.~\ref{fig: anomal }(a). The wavefront changes the direction, and the outgoing SWs are bent. The arrows in Fig.~\ref{fig: anomal }(a) indicate the direction of the refracted waves estimated with Eq.~(\ref{eq: snell}). Although the gradient of the phase shift is not perfectly linear, this estimation seems to be very good.

The results of the numerical simulations are also compared to the results of the analytical consideration using the Huygens--Fresnel principle.
Complex amplitudes at every position are calculated according to the postulation given by Eq.~(\ref{eq: hug}).
We assume that at the position $d$ (as it is shown in  Fig.~\ref{fig: anomal }(c)), we have a line of sources of cylindrical waves. Fig.~\ref{fig: msani}(a) presents the transmittance and phase shift of the SW that we need to put into Eq.~(\ref{eq: hug}). The application of this equation is shown in Fig.~\ref{fig: anomal }(c).
Since the phase shift changes smoothly along the $y$-axis, the wavefront of the SWs transmitted through the slab is reconstructed in the way that we observe the bending of SWs of $\sim$36$^\circ$. We can conclude that the procedure based on the Huygens--Fresnel principle and the approach based on the generalized Snell's law are suitable for the estimation of the results of the numerical simulation, and these analytical approaches can describe the system.

\subsection{Graded index slab at the bent of the magnonic waveguide}

We will use the GRIN slab, designed in the previous section, to guide the SWs through the bent of the waveguide. The numerical study will be performed with the aid of micromagnetic simulations (see Appendix B).

\begin{figure}

{\includegraphics[width=1\columnwidth]{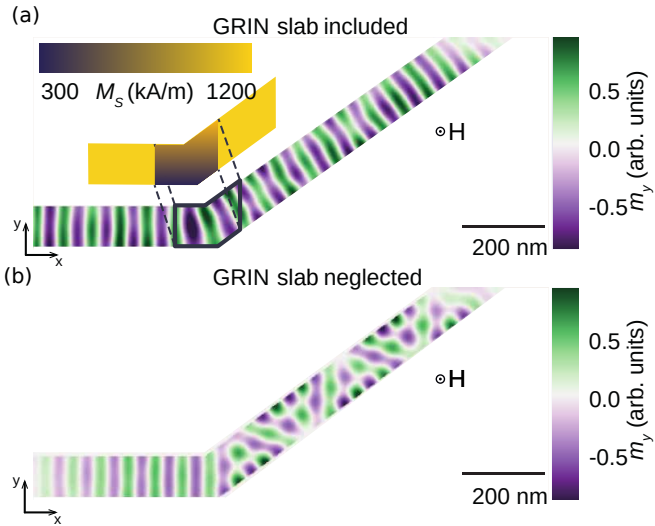}
\caption{\label{fig:full+empty}(a) The SWs propagation in the CoFeB waveguide, where the GRIN slab is placed at the bent, releases the anomalous refraction. The GRIN slab has the same gradient of $M_\mathrm{S}$ as it was used in Fig.~\ref{fig: anomal }. The color map represents a dynamical component of the magnetization in the $y$ direction. A snapshot is taken at the moment when a steady state is reached. At the beginning of the horizontal branch of the waveguide, the microwave antenna excites the SWs and at the end of the tilted branch, the SWs are damped to avoid any reflection. After taking a turn, the SWs propagate smoothly.
(b) The SWs propagation in the waveguide without a gradient of magnetic parameters. After taking a
turn, the SWs show a complex behavior due to interference between different modes of the waveguide. The width of the waveguide is $100$ nm. The material parameters are the same as for the system presented in Fig.~\ref{fig: anomal }.
}
}
\end{figure}

Let us consider two straight sections of the flat waveguides of the width $100$ nm, made of CoFeB, and connected at the angle $36^{\circ}$. At this angle, we observe the refraction of the SWs by the GRIN slab with the assumed gradient of the saturation magnetization  (Fig.~\ref{fig: anomal }). In Fig.~\ref{fig:full+empty}(a), we present this curved waveguide with such a GRIN slab placed at the bent, whereas the results for SWs propagation through the same structure without the GRIN element are presented in Fig.~\ref{fig:full+empty}(b). These two systems guide the SW's differently. We can see the SWs' interference pattern in the outgoing section of the bent waveguide without the GRIN slab. Such behavior results from the scattering of the incident fundamental mode (not quantized across the waveguide's width) to the higher modes (quantized across the waveguide's width).
As a consequence, the information encoded into the phase of the incident fundamental mode is lost. On the other hand, the application of the GRIN slab introduces the anomalous refraction at waveguide's bent causing that
the outgoing waves propagate along the waveguide in the form of the fundamental mode with non-disturbed wavefronts and well-defined phase.

In the previous calculations, the dipolar interactions were neglected, since their full consideration in the boundary condition problem is a complicated task. 
Therefore, all investigations were limited to the exchange SWs, including the micromagnetic simulations. 
Such approach is justified because the exchange interactions dominate over the dipolar ones for short wavelength SWs. However, in order to verify the applicability of the GRIN slab, we performed simulations that also included dipolar interactions. 

\begin{figure}
{\includegraphics[width=1\columnwidth]{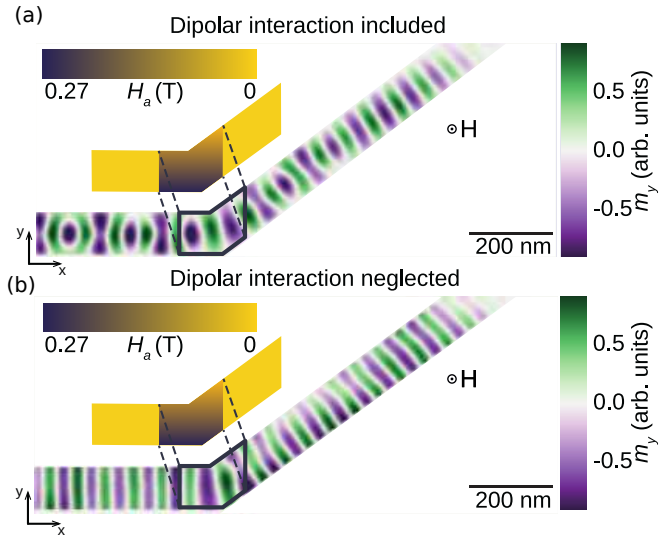}
\caption{\label{fig: ani+demag} The SW propagation in the bent waveguide of the same geometry, as in Fig.~~\ref{fig:full+empty}. The GRIN slab placed at the bend is characterized by the gradient of anisotropy field $H_\mathrm{a}$. The gradient of $H_\mathrm{a}$ was chosen to obtain refraction at the same angle, as in Fig.~\ref{fig:full+empty}. In panel (a) the dipolar interaction is included and in panel (b) it is neglected in the micromagnetic simulations.
}
}
\end{figure}

Let us discuss the waveguide with a GRIN slab in which the gradient of the anisotropy field is introduced. In this scenario, the static dipolar magnetic field is uniform throughout the whole system, i.e., if  $H>M_\mathrm{S}$, like in our system, $\mathbf{H}_{\mathrm{dem,0}}=-\hat{\mathbf{z}}M_\mathrm{S}$, since the saturation magnetization is uniform, what makes the system easier to model. 
Following Fig.~\ref{fig: msani}(b), we chose the range of anisotropy, which needs to be applied to have the same bending of SWs, as for the studies presented in Fig.~\ref{fig:full+empty}. Keeping the same shape of the GRIN slab, we chose the gradient of the uniaxial anisotropy field in the range from $0.27$~T to $0$. 
The results are compared with the case when the dipolar interactions are included in the micromagnetic simulations, for the same geometry and the values of magnetic parameters.
For the 100-nm-wide and 5-nm-thick waveguide, the static demagnetizing field shifts down the dispersion relation, so in order to keep the same wave vector for the considered frequency ($25$~GHz), it is necessary to apply the external magnetic field of increased value with respect to simulations without the dipolar interactions. 
We analyzed the dispersion relations and found out, that an additional field  $\mu_0 H=1.35$ T compensates the effect of the static demagnetizing field. 
The cases when the dipolar interaction is taken into account or neglected, are compared in Fig.~\ref{fig: ani+demag}. Only a small difference is visible, which means that the impact of the dipolar field is not significant. 
The difference results from different boundary conditions at the edges of waveguides, where spins are partially pinned\cite{Guslienko2002bc}.

It is necessary to note, that in the case of the GRIN slab with the gradient of saturation magnetization in the out-of-plane magnetized thin film, the static demagnetizing field is not uniform. As a result, both the saturation magnetization and static effective magnetic field are changed in parallel, and it also needs to be taken into account in the design of the GRIN slab for SW steering, which is out of the scope of this paper. 

\bigbreak
\bigbreak
\section{Summary}
 We presented a comprehensive study of the SWs propagation through the GRIN slab with spatially modulated magnetization saturation or uniaxial anisotropy. 
 Using the analytical model, crosschecked by numerical simulations, we were able to relate the phase acquired by SWs during the transmission through the GRIN slab at different locations to the values of the spatially dependent magnetic parameters. 
 With this knowledge, we designed the GRIN slab, ensuring the phase-coherent refraction of SWs at a desirable angle, determined by the gradient of saturation magnetization or the gradient of the anisotropy field. Notably, the presented GRIN slab enabled a wide range of phase changes (ca. $1.75\pi$) for high and uniform transmission of SWs (transmittance in the range of 80\%-100\%) enhanced by utilizing the Fabry-Perot resonances. The analytical model and the main numerical demonstrations were performed for the exchange SWs, later validated with micormagnetic simulations for selected cases with included dipolar interactions.
 
 As an application, we demonstrated GRIN slab used to guide the SWs coherently in fundamental mode along the bend in a magnonic waveguide. The micromagnetic simulations were performed for realistic structure and showed the possibility for experimental realization of this idea. 

The GRIN slab considered in this paper can be activated on demand by the voltage induced anisotropy\cite{Rana19}. The spatial changes of the perpendicular anisotropy can be introduced by the variation of the thickness of insulating spacer separating one of the electrodes. This can open the route for voltage-controlled routing of SWs.

 The proposed approach can be verified experimentally by the state of the art techniques, like an XMCD\cite{Grafe2019, Lisiecki2019}, phase-resolved BLS\cite{Vogt2009} or broadband microwave spectroscopy\cite{Baumgaertl2018, Dobrovolskiy19}.
 We believe that our findings are substantial for further development of the circuits for analog and digital computing based on SWs and contribute to the field of magnonics.

\section{Acknowledgments}

{$\dagger$}Szymon Mieszczak and Oksana Busel contributed equally to this work.

The study has received financial support from the National Science
Centre of Poland - Grant No.: UMO-2016/21/B/ST3/00452 and support from the EU -
Horizon 2020 project MagIC Grant No.: 644348. 
\appendix
\section{Spatial profiles of the spin waves, resonantly scattered at the magnetic slab}

Solution of the LLE, Eq.~(\ref{Eq:LLE}), provides the information about the complex amplitudes of SWs
which propagate in the system. The one-dimensional model presented in this paper has the following solutions:
\begin{equation}
t_{\mathrm{A}}=\frac{4e^{id(k_{\mathrm{B}}-k_{\mathrm{A}})}}{\alpha_{\mathrm{A}}\alpha_{\mathrm{B}}k_{\mathrm{A}}k_{\mathrm{B}}\left(f^{+}f^{+}-e^{2idk_{\mathrm{B}}}f^{-}f^{-}\right)},\label{eq:t_A}
\end{equation}
\begin{equation}
r_{\mathrm{A}}=\frac{-f^{-*}f^{+}+e^{2idk_{\mathrm{B}}}f^{+*}f^{-}}{\left(f^{+}f^{+}-e^{2idk_{\mathrm{B}}}f^{-}f^{-}\right)},\label{eq:r_A-1}
\end{equation}
\begin{equation}
t_{\mathrm{B}}=\frac{2f^{+}}{\text{\ensuremath{\alpha_{\mathrm{B}}}}k_{\mathrm{B}}\left(f^{+}f_{\mathrm{BA}}^{+}-e^{2idk_{\mathrm{B}}}f^{-}f^{-}\right)},\label{eq:t_B-1}
\end{equation}
\begin{equation}
r_{\mathrm{B}}=\frac{2e^{2idk_{\mathrm{B}}}f^{-}}{\text{\ensuremath{\alpha_{\mathrm{B}}}}k_{\mathrm{B}}\left(f^{+}f^{+}-e^{2idk_{\mathrm{B}}}f^{-}f^{-}\right)},\label{eq:r_B-1}
\end{equation}
where $f^{\pm}$ is an auxiliary function, namely:
\[
f^{\pm}=\frac{M_{\mathrm{B}}}{\alpha_{\mathrm{A}}M_{\mathrm{A}}k_{\mathrm{A}}}\pm\frac{M_{\mathrm{A}}}{\alpha_{\mathrm{B}}M_{\mathrm{B}}k_{\mathrm{B}}}-\frac{i}{A},
\]
and $f^{\pm*}$ is its complex conjugate.
The dependence of the wave number on angular frequency 
$k_{\mathrm{A}\left(\mathrm{B}\right)}(\omega)$ expresses the dispersion relation:
\begin{equation}
k_{\mathrm{A}\left(\mathrm{B}\right)}=\frac{1}{\sqrt{\alpha_{\mathrm{A}\left(\mathrm{B}\right)}M_{S,A\left(\mathrm{B}\right)}}}\sqrt{\frac{\omega}{\gamma}-\mu_0\left(H+H_{\rm{a},{\mathrm{A}\left(\mathrm{B}\right)}}\right)},\label{eq:k_A(B)}
\end{equation}
where $H$ and $H_{\rm{a},{\mathrm{A}\left(\mathrm{B}\right)}}$ are external and anisotropy fields, respectively.

Parameter $\delta$  is the thickness of the interfaces and it is a fitting parameter. 
We found out that the best match between analytical and numerical results is achieved for the value of $\delta=0.5$~nm. It is reasonable value since the size of the unit cell in micromagnetic simulations is 1 nm. 

Information about the energy flow (the amplitude of SWs) and phase shifts can be extracted from Eqs.~(\ref{eq:t_A})--(\ref{eq:r_B-1}), and as we can see in Figs. \ref{fig:freqms}--\ref{fig: msani}, both quantities fit perfectly to micromagnetic simulations.

Let us present the spatial profile of dynamical components of the magnetization vector.
Set of Eqs.~(\ref{eq:mA1})--(\ref{eq:mA2}), with coefficients defined in Eqs.~(\ref{eq:t_A})--(\ref{eq:r_B-1}), 
describe the analytical solution of SW in complex form. To visualize these spatial profiles, the magnetization should be presented in the real form:
\begin{equation}
    m = \Re\left[ \tilde{m}e^{i\left(kx-\omega t\right)}\right] =\left|\tilde{m}\right|\cos\left(kx-\omega t+\arg \tilde{m}\right),
\end{equation}
where $\tilde{m}$ is a complex coefficient defined for specific region. It can be incidence, reflection or transmission coefficient, as defined in  Eqs.~(\ref{eq:t_A})--(\ref{eq:r_B-1}). 

\begin{figure}
{\includegraphics[width=1\columnwidth]{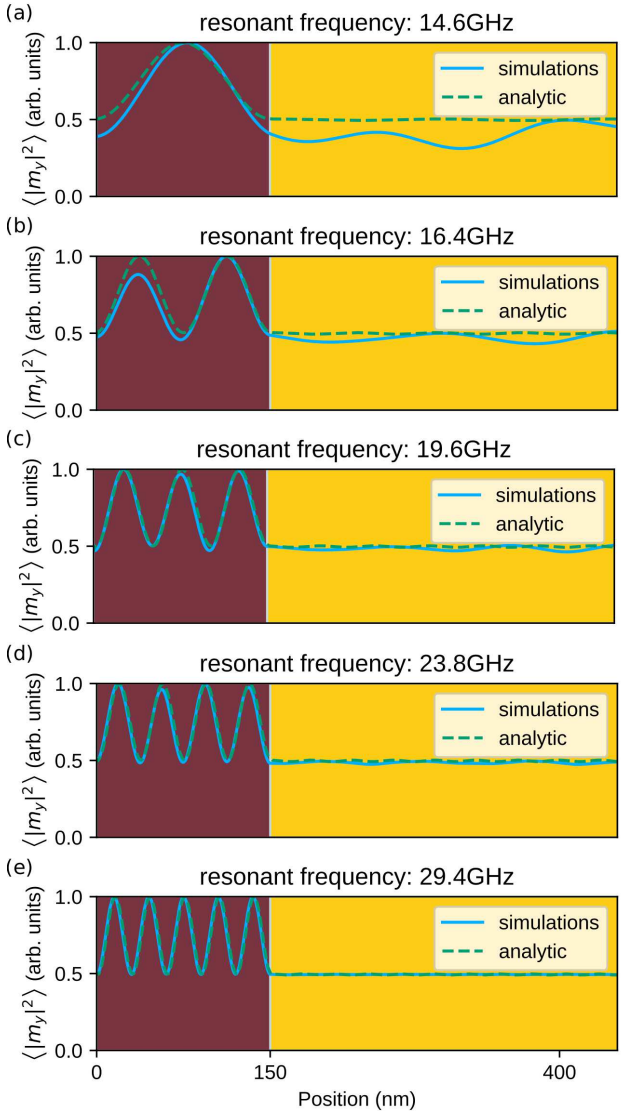}
\caption{\textcolor{red}{\label{fig:resonances}}The normalized spatial profile
of dynamical component of the magnetization for the five lowest resonances
(a) $14.6$~GHz, (b) $16.4$~GHz, (c) $19.6$~GHz, (d) $23.8$~GHz and (e) $29.4$~GHz.
Width of the slab is $150$~nm, $M_{\rm{S}}$ in the matrix is $1200$~kA/m
and $A_{\rm{ex}}=28$~pJ/m. External field is $0.5$~T, reduced $M_{\rm{S}}$
within the slab is $800$~kA/m and reduced exchange stiffness constant $A_{\text{ex}}$ is $20$~pJ/m.
The presented resonances correspond to Fig.~\ref{fig:freqms}.
}
}
\end{figure}

In order to compare the results of the analytical model with the ones from numerical simulations, we need to get rid of the explicit dependence on time, because we are not able to compare some exact moments in time. Hence, we average the squared magnetization component in time. The same we did with the output from simulation after reaching a steady state. 
The comparison is shown in Fig.~\ref{fig:resonances}. We plotted the spatial profile of SW for five specific frequencies (see Fig.~\ref{fig:freqms}), Fig.~\ref{fig:resonances}(a)--(e), where transmittance is equal to 1 (see Fig.~\ref{fig:freqms}). 
Dark background represents the slab B. Standing wave is visible, that indicates the existence of resonant effect in accordance with Eq.~(\ref{eq: resonance}). The slab is coupled to the surrounding, and the transmission is relatively high. Therefore, the nodes are not located at zero level, because the amplitude of the reflected wave is always lower than the amplitude of the incident wave.
On the right areas (light background), SWs are propagating with a constant amplitude,
so after averaging, we get a straight line. Its level indicates the energy
flow. 

Fig.~\ref{fig:resonances} is another remarkable confirmation of the validity of our analytical model. 

\section{Micromagnetic simulations}

The micromagnetic simulations were performed with the use of mumax3 package \cite{Vansteenkiste2014}, which is the finite-difference time-domain solver of full Landau-Lifshitz equation.
The simulations were conducted to (i) calculate the transmittance and the phase shift of the waves transmitted through the one-dimensional slab, (ii) demonstrate anomalous refraction in thin-film, and finally, (iii) to demonstrate a slab suitable for SW bending in magnonic waveguides. 
In all these cases, the steady state were simulated, i.e., the response of the system after long, continuous excitation of SWs at a single frequency enabling SWs to travel to the sides of the simulated system.

In order to neglect the influence of the reflections from the borders of the simulated system and, therefore, simplify interpretation of the results, the absorbing boundary conditions are implemented around the sides of the simulated domain (see Fig. \ref{fig: geometry_of_the_system_2}). In all the simulations we consider SW propagation in 5-nm-thick CoFeB film ($M_\mathrm{S}=1200$ kA/m, $A_\mathrm{ex}=27$ pJ/m and reduced damping to 0.0001) in the presence of the out-of-plane magnetic field (applied along the $z$-axis) of value $\mu_0 H=0.5$ T and discretized with cuboid elements of dimensions (1 nm$\times$1 nm$\times$5 nm). These values are comparable to the exchange length. Magnetic parameters were modulated only in the region of the slab as it is described in the main part of the paper. In order to be consistent with the analytical theory, which neglects the dipolar interactions, most of the simulations were performed with neglected dipolar interactions as well. Nevertheless, in order to further validate this model, a set of micromagnetic simulations with included dipolar interaction was performed to demonstrate the applicability of the considered slabs to bend SW in a curved waveguide (see Fig.~\ref{fig: ani+demag}). 

To calculate the transmittance and the phase shift of the transmitted SW through the slab with respect to frequency or various magnetic parameters, we defined a one-dimensional geometry, i.e., discretized by $L_{x}\times L_{y}\times L_{z} \left(4096\times1\times1\right)$ unit cells. These values were obtained by running two separate simulations, i.e., a reference simulation without the slab, and the additional one with the slab. Then, the results of these simulations were compared in order to extract the values of transmittance and phase shift. The transmittance was calculated as a ratio of squared amplitude and the phase shift -- as a difference of the SWs phases. SWs were excited by an RF magnetic field of frequencies in the range of $14-40$ GHz for fixed material parameters or $25$ GHz for various material parameters of the slab. RF field is applied locally in a $6$ nm narrow region, located on the left side of the slab at a distance of $1548$ nm from the slab.

To demonstrate an anomalous refraction in action, we defined a two-dimensional system discretized by $L_{x}\times L_{y}\times L_{z} \left(1024\times1024\times1\right)$ unit cells. The scheme of the system is presented in Fig.~\ref{fig: geometry_of_the_system_2}. SWs are excited by an RF field of frequency $25$ GHz applied locally in a $6$ nm narrow region located on the left side of the slab in distance of $232$ nm. 
Three different areas can be distinguished there. On the left is the narrow part, $100$ nm-wide waveguide, which introduces SWs into the system. In the middle is a 100-nm-wide and 150-nm-long slab with a gradient of the magnetic parameters. 
The gradient of the magnetic parameters is induced along the $y$-axis.
Saturation magnetization changes in the range of 300-800 kA/m on the distance $100$ nm, with direct contact with waveguide. Above is smoothly changed in the range of 800-1200 kA/m to avoid sharp edges. Below, a constant value of 300 kA/m is assumed. On the right-hand side is the semi-infinite medium, which allows the propagation of SWs in any direction freely. 

Finally, using the knowledge from the previous step, we design a curved waveguide, which supports a coherent SWs propagation alongside the waveguide. To create such a curved waveguide, we cut the finite 2D system following the newly created wavefront. The system is discretized by $L_{x}\times L_{y}\times L_{z} \left(2048\times512\times1\right)$ unit cells. SWs are excited by an RF magnetic field of frequency $25$ GHz applied locally in $6$ nm narrow region located at the distance of $224$ nm to the corner from the left side.

\bibliography{bibliography}

\end{document}